\documentclass[journal=jpccck,manuscript=article]{achemso}

\usepackage{chemformula} 
\usepackage[T1]{fontenc} 

\author{Shuhang Chen}
\affiliation{College of Science, University of Shanghai for Science and Technology, Shanghai 200093, China}
\author{Wenjing Xu}
\affiliation{College of Science, University of Shanghai for Science and Technology, Shanghai 200093, China}
\author{Yueyue Ning}
\affiliation{College of Science, University of Shanghai for Science and Technology, Shanghai 200093, China}
\author{Ke Yang}
\email{keyang@usst.edu.cn}
\affiliation{College of Science, University of Shanghai for Science and Technology, Shanghai 200093, China}

\title[An \textsf{achemso} demo]
{Exploration of the two-dimensional Ising magnetic materials in the triangular prismatic crystal field}

\begin{document}

\begin{tocentry}


%
\includegraphics[width=8.2cm]{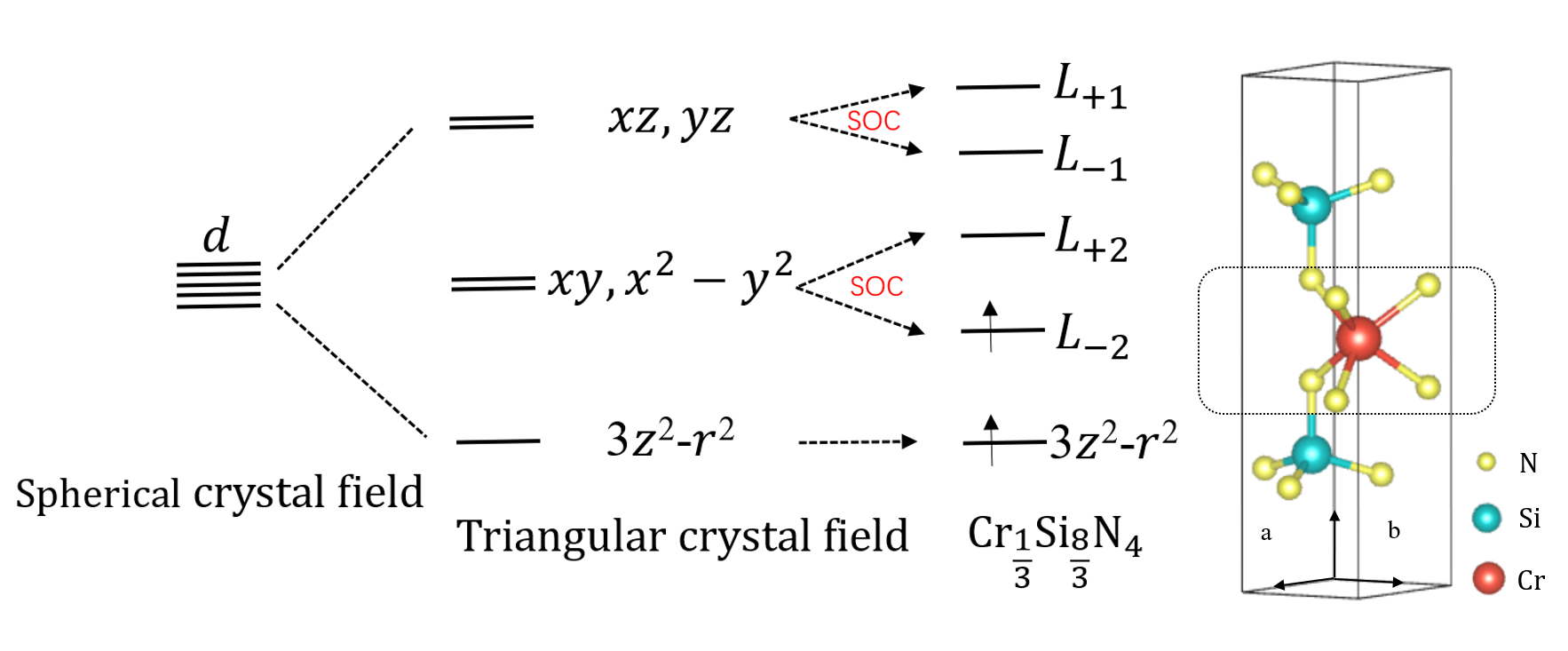}
%


\end{tocentry}

\begin{abstract}
Magnetic anisotropy is essential for stabilizing two-dimensional (2D) magnetism, which has significant applications in spintronics and the advancement of fundamental physics.
In this work, we examine the electronic structure and magnetic properties of triangular prismatic MSi$_2$N$_4$ (M = V, Cr) monolayers, using crystal field theory, spin-orbital state analyses,  and density functional calculations. 
Our results reveal that the pristine VSi$_2$N$_4$ monolayer exhibits magnetism with a V$^{4+}$ 3$d^1$ $S$ = 1/2 charge-spin state within the triangular prismatic crystal field. However, the strong $d$ orbital hybridization between adjacent V$^{4+}$ ions disrupts the $d$ orbital splitting in this crystal field, resulting in a relatively small in-plane magnetic anisotropy of approximately 2 $\mu$eV per V atom.
In contrast, the pristine CrSi$_2$N$_4$ monolayer is nonmagnetic, characterized by the Cr$^{4+}$ 3$d^2$ $S$ = 0 state. Upon substituting nonmagnetic Cr$^{4+}$ with Si$^{4+}$, Cr$_\frac{1}{3}$Si$_\frac{8}{3}$N$_4$ transforms into an antiferromagnetic insulator with Cr$^{4+}$ 3$d^2$ $S$ = 1 state, featuring a large orbital moment of --1.06 $\mu_{\rm B}$ oriented along the $z$-axis and huge perpendicular magnetic anisotropy of 18.63 meV per Cr atom.
These findings highlight the potential for further exploration of 2D Ising magnetic materials within a unique triangular prismatic crystal field.
\end{abstract}

\section{Introduction}
Since the discovery of two-dimensional (2D) ferromagnetism (FM) in atomically thin layers CrI$_3$\cite{Huang2017} and Cr$_2$Ge$_2$Te$_6$\cite{Gong2017}, two-dimensional magnetic materials have been the subject of extensive research due to their unique electronic structure and rich tunable properties\cite{Gibertini2019,Gong2019,Song2019,Burch2018,Mak2019}. 
The Mermin-Wagner theorem establishes that 2D isotropic Heisenberg spin systems lack long-range magnetic order at any non-zero temperature\cite{Mermin1966}.  As a result, magnetic anisotropy (MA) becomes critical for stabilizing 2D magnetic systems. Moreover, a huge MA offers greater resistance to thermal perturbations, which is beneficial for ensuring stable data storage and enhancing the reliability of spintronic devices\cite{Huang2017, Gong2017}.
Both the CrI$_3$ monolayer\cite{Huang2017} and Cr$_2$Ge$_2$Te$_6$\cite{Gong2017} have a closed $t_{2g}^3$ shell for the octahedral Cr$^{3+}$ $S$ = 3/2 ion. As a result, no single-ion anisotropy (SIA) is produced from the orbital singlet. Instead, the finite perpendicular MA in these materials arises from exchange anisotropy, which is induced by spin-orbit coupling (SOC) involving the ligand heavy $p$ orbitals and their hybridization with Cr 3$d$ orbitals\cite{Lado2017,Kim2019}.
The search for materials with huge MA, especially 2D Ising magnetic materials, is pivotal for the advancement of spintronics. Their resilience to thermal perturbations enables their use in room-temperature magnetic applications at the 2D limit. This breakthrough is key in developing next-generation spintronic devices, promising to revolutionize information storage and processing with improved efficiency, speed, and miniaturization~\cite{Lado2017,Kim2019,Xu2018,Yang2020,Liu2020,Yang2021,Sears2020,Ni2021}.

\begin{figure}[t]
	\centering
	\includegraphics[width=8cm]{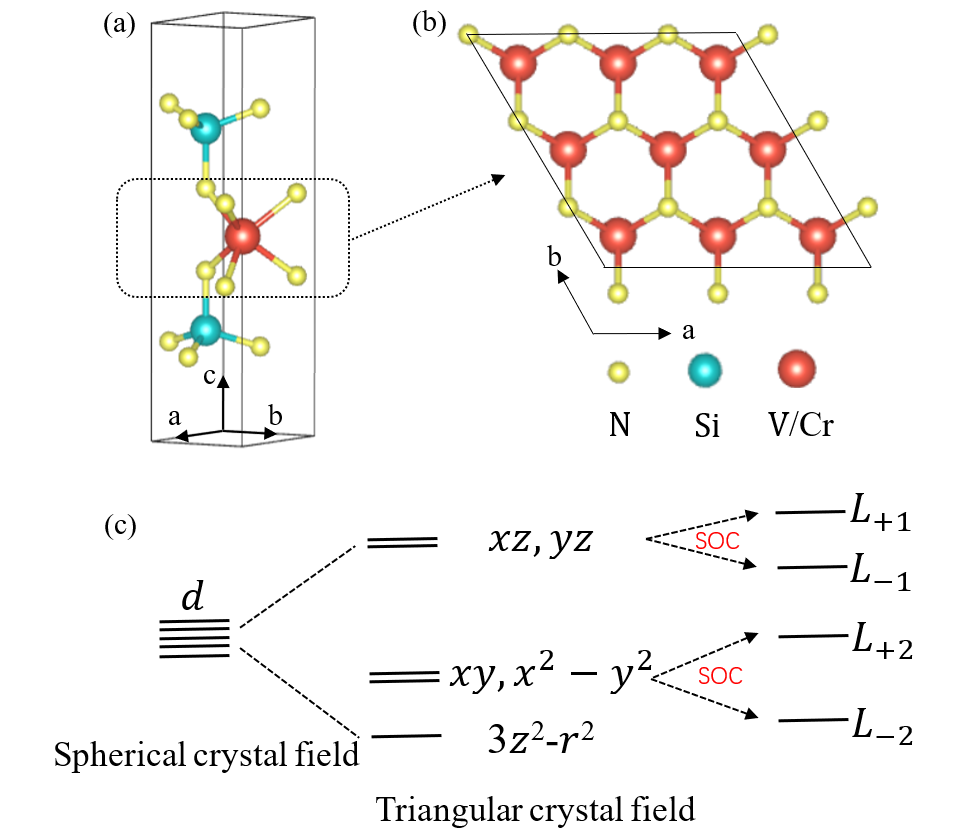}
	\centering
	\caption{(a) The crystal structure of the VSi$_2$N$_4$ (CrSi$_2$N$_4$) monolayer, which can be viewed as a VN$_2$ (CrN$_2$) layer sandwiched between two Si-N layers. (b) The top view of the $3\times3\times1$ structure for VN$_2$ (CrN$_2$) layer with the surrounding Si-N layers omitted for clarity. (c) The 3$d$ orbitals crystal field in the unique triangular prismatic crystal field. 
	}
	\label{Fig.1}
\end{figure}

Recently, MSi$_2$N$_4$ monolayer, where M represents either Mo or W, has been successfully synthesized using chemical vapor deposition techniques\cite{Hong2020}. These materials show semiconductor characteristics and are notable for their excellent electronic properties as well as their robust environmental stability.
Building on this achievement, there has been heightened research interest in the MX$_2$Y$_4$ family of layered materials~\cite{Ren2022,Tho2023}. Noteworthy research includes the discovery of unique electronic effects in CrSi$_2$Y$_4$ (Y = N, P)\cite{Liu2021}, the electronic and excitonic properties in MSi$_2$Y$_4$ (M = Mo, W; Y = N, P, As, Sb) films\cite{Wozniak2022}, the exploration of new stable Janus MX$_2$Y$_4$ (M = Sc$\sim$Zn, Y$\sim$Ag, Hf$\sim$Au; X = Si, Ge; Y = N, P) monolayers for potential applications\cite{Zhang2022}, the investigation of the electronic structure of MX$_2$Y$_4$ (M = Ti, Cr, Mo; X = Si; Y = N, P) bilayers under vertical strain\cite{Zhong2021}, an approach to construct MX$_2$Y$_4$ family monolayers with a septuple-atomic-layer structure\cite{Wang2021}, and the prediction of an indirect band-gap semiconductor Janus MSiGeN$_4$ (M = Mo, W) monolayers\cite{Guo2021}.

In the case of MSi$_2$N$_4$, where M adopts a +4 charge state within a unique triangular prismatic crystal environment\cite{Hong2020}. The five $d$ orbitals split into three distinct sets: a lowest-energy singlet labeled $3z^2-r^2$, a middle doublet denoted as $xy$/$x^2-y^2$, and the highest-energy doublet, $xz$/$yz$, as shown in Fig. 1(c). When the SOC is considered, these initially degenerate doublets ($xy$/$x^2-y^2$ and $xz$/$yz$) further separate into complex $L_{\pm2}$ and $L_{\pm1}$ orbitals, respectively. This further orbital splitting, induced by SOC, leads to a huge perpendicular SIA\cite{Lado2017,Kim2019}. Therefore, the unique triangular prismatic crystal field in these MSi$_2$N$_4$ monolayers is conducive to achieving huge MA.

In this study, we employ crystal field theory, spin-orbital state analyses, and density functional calculations to examine the MSi$_2$N$_4$ (M = V, Cr) monolayers. Our findings indicate that the transition metals (M) exist in a +4 valence state within the triangular prismatic crystal field. However, the strong hybridization between neighboring M$^{4+}$ ions interferes with the $d$ orbital splitting, leading to distinct magnetic properties. Specifically, pristine VSi$_2$N$_4$ exhibits a magnetic state characterized by the V$^{4+}$ ion with a $S$ = 1/2 charge-spin state. 
Our findings reveal that VSi$_2$N$_4$ features an orbital singlet, which results in a small in-plane MA of approximately 2 $\mu$eV per V atom.
In contrast, the pristine CrSi$_2$N$_4$ monolayer is nonmagnetic, characterized by the Cr$^{4+}$ 3$d^2$ $S$ = 0 state. However, when the nonmagnetic Cr$^{4+}$ is substituted by Si$^{4+}$, the Cr$_\frac{1}{3}$Si$_\frac{8}{3}$N$_4$ monolayer transforms into an antiferromagnetic insulator with Cr$^{4+}$ 3$d^2$ $S$ = 1 state. Moreover, the Cr$^{4+}$ ions display a large orbital moment of --1.06 $\mu_B$, oriented along the $z$-axis, as well as a huge perpendicular MA energy of 18.63 meV per Cr atom in the Cr$_\frac{1}{3}$Si$_\frac{8}{3}$N$_4$ monolayer. 
Therefore, we demonstrate that the triangular prismatic crystal field serves as a promising experimental platform for discovering novel 2D Ising materials.

\section{Methods}
The density functional theory (DFT) calculations are carried out using the Vienna ab initio Simulation Package (VASP)~\cite{Kresse1996}. The exchange-correlation effect is described by the generalized gradient approximation (GGA) using the functional proposed by Perdew, Burk, and Ernzerhof (PBE)~\cite{Perdew1996}. The kinetic energy cutoff for plane-wave expansion is set to 450 eV.
A $3\times3\times1$ supercell is chosen for MSi$_2$N$_4$.
The crystal structure of the VSi$_2$N$_4$ (CrSi$_2$N$_4$) monolayer is illustrated in Fig. \ref{Fig.1}(a). It can be understood as the incorporation of a 2H phase VN$_2$ (CrN$_2$) layer into an $\alpha$-InSe-type Si$_2$N$_2$ framework. The resulting composite forms a hexagonal lattice with the $P\overline{6}m2$ space group.

The local V ion has a triangular prismatic crystal field, splitting the degenerate five $d$ orbitals into the lowest $3z^2-r^2$ singlet, middle $xy$/$x^2-y^2$ doublet and the highest $xz$/$yz$ doublet. Taking into account the SOC, the degenerate $x^2-y^2$/$xy$ ($xz$/$yz$) doublet can further split into $L_{\pm2}$ ($L_{\pm1}$) orbitals (see Fig. \ref{Fig.1} (c)). The complex orbital wave functions can be written as
\begin{equation} \label{eq: 1}
	\begin{aligned}
		&\ L_{\pm2}=\frac{1}{\sqrt{2}}(d_{x^{2}-y^{2}}\pm id_{xy})&\\
		&\ L_{\pm1}=\frac{1}{\sqrt{2}}(d_{yz}\mp id_{xz})&\\
	\end{aligned}
\end{equation}
Thus, the utilization of the triangular prismatic crystal field emerges as a promising avenue for the achieving of 2D Ising magnetization with huge MA.

To better describe the on-site Coulomb interactions of V (Cr) 3$d$ electrons, the typical value of the Hubbard $U_{\rm eff}$ = 4.0 eV is used in the GGA+$U$ calculations~\cite{Anisimov1993}. The spin-orbit coupling (SOC) is also included in our GGA+SOC+$U$ calculations to study MA.
The method for controlling the density matrix is implemented using open-source software developed by Watson~\cite{Allen2014}, and this method has been widely used in previous studies~\cite{He2021,Yang2020,Dorado2009,Ou2014,Varignon2019}. In addition, using the site-projected wave function character of $d$ orbitals in DFT calculations, we can project the density of states (DOS) based on different eigen wavefunctions, which are linear combinations of the five common $d$ orbitals ($xz$, $yz$, $xy$, $x^2-y^2$, $3z^2-r^2$).
As seen in equation (\ref{eq: 1}), this method can be used in the complex crystal field and the complex orbitals due to SOC, which is essential in our calculations.

\section{Results \& Discussion}
\begin{figure}[H]
	\includegraphics[width=8cm]{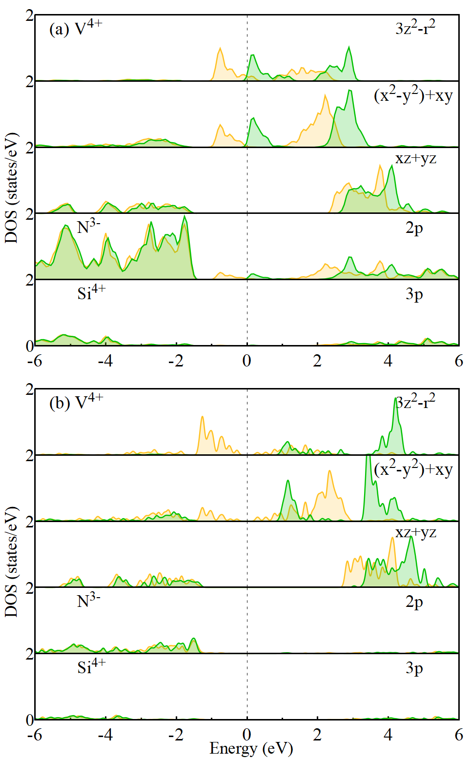}
	\centering
	\caption{The FM VSi$_2$N$_4$ DOS results of the V$^{4+}$ 3$d$, N$^{3-}$ 2$p$ state and Si$^{4+}$ 3$p$ state within (a) GGA+spin and (b) GGA+$U$+SOC framework. The orange (green) curve stands for the up (down) spin. The Fermi level is set at zero energy.
	}
	\label{Fig.2}
\end{figure}

We first perform the spin-polarized GGA calculations to study the electronic and magnetic structure of the VSi$_2$N$_4$ monolayer. In the FM state, the calculated V local spin moment of 0.86 $\mu_{\rm B}$ refers to the nominal V$^{4+}$ 3$d^1$ $S$ = 1/2 state. The reduction of V$^{4+}$ spin moment is due to the strong covalency with the nitrogen ligands. The total spin moment of 0.92 $\mu_{\rm B}$ per V atom agrees well with the V$^{4+}$, Si$^{4+}$ and N$^{3-}$ charge states in the VSi$_2$N$_4$ monolayer. We plot in Fig. \ref{Fig.2}(a) the orbitally resolved DOS for the FM state. We find that the $xz$/$yz$ doublet has the highest energy about 3 eV above the Fermi level. In contrast, the lowest $3z^2-r^2$ singlet and middle $x^2-y^2$/$xy$ doublet have a strong hybridization, both of which are on average 1/3 filled to fulfill the formal V$^{4+}$ 3$d^1$ state. Due to the strong V-N covalency in the VN$_2$ layer, the adjoining nitrogen becomes negatively spin-polarized and has a local spin moment of --0.04 $\mu_{\rm B}$. The Si$_2$N$_2$ layer is not affected by spin-polarized V$^{4+}$ ions and remains nonmagnetic.

\renewcommand\arraystretch{1.3}
\begin{table}[t]
	\caption{Relative total energies $\Delta$\textit{E} (meV per V atom), total spin moment ($\mu_{\rm B}$ per V atom), local spin moment ($\mu_{\rm B}$), and orbital moment ($\mu_{\rm B}$) of FM state for VSi$_2$N$_4$, V$_\frac{1}{9}$Si$_\frac{26}{9}$N$_4$ ($3z^2-r^2$ and $L_{-2}$ states) and V$_\frac{1}{3}$Si$_\frac{8}{3}$N$_4$ ($3z^2-r^2$ and $L_{-2}$ states). The ${\perp}$ and ${\parallel}$ stand for the out-of-plane magnetization and in-plane magnetization. 
	}
	\begin{tabular}{c@{\hskip2mm}c@{\hskip2mm}c@{\hskip2mm}c@{\hskip2mm}c@{\hskip2mm}c@{\hskip2mm}}
		\hline\hline
		Systems & States & $\Delta$\textit{E} & V$_{\rm spin}$ & \textit{M}$_{\rm tot}$  & V$_{\rm orb}$ \\ \hline
		{VSi$_2$N$_4$}&$\perp$& 0 & 1.12       & 1.00   & --0.01    \\
		&$\parallel$ & --0.002 & 1.12       & 1.00   & --0.02    \\\hline
		{V$_\frac{1}{9}$Si$_\frac{26}{9}$N$_4$}  &  $3z^2-r^2$, $\perp$ & 0 & 1.08       &   1.00 & --0.01    \\
		&$3z^2-r^2$, $\parallel$ & --0.059 & 1.08       & 1.00  & --0.02    \\
		&$L$=--2, $\perp$ & 755.636 (0) & 0.98  & 1.00 & --1.32\\
		&$L$=--2, $\parallel$ & 778.157 (22.521) & 0.99    & 1.00 & --1.32  \\
		\hline
		{V$_\frac{1}{3}$Si$_\frac{8}{3}$N$_4$}  &  $3z^2-r^2$, $\perp$ & 0 & 1.09       &  1.00  & --0.01    \\
		&$3z^2-r^2$, $\parallel$ & --0.001 & 1.09       &  1.00  & --0.02    \\
		&$L$=--2, $\perp$ & 742.098 (0) & 1.00  & 1.00  & --1.32\\
		&$L$=--2, $\parallel$ & 754.344 (12.246) & 0.86    & 0.85 & --1.31  \\
		\hline\hline
	\end{tabular}
	\label{tb1}
\end{table}

To account for electronic correlations of the V ions and to incorporate SOC effect, we carry out the GGA+SOC+$U$ calculations with the out-of-plane magnetization axis. As shown in Table \ref{tb1}, the enhanced electron localization gives rise to an increasing local spin moment of 1.12 $\mu_{\rm B}$ for V ion, and a small orbital moment of --0.01 $\mu_{\rm B}$ induced by SOC.
The total spin moment of 1.00 $\mu_{\rm B}$ per V atom suggests the formal V$^{4+}$ 3$d^1$ $S$ = 1/2 state again. A clear 0.2 eV semiconductor energy gap is observed, as seen in Fig. \ref{Fig.2}(b). The electron of V$^{4+}$ ion mainly occupies the lowest $3z^2-r^2$ singlet, while the higher $x^2-y^2$/$xy$ doublet remains partially occupied. This indicates that 
the $3z^2-r^2$ and $x^2-y^2$/$xy$ orbitals have a strong hybridization even in localized V ions, and the phenomenon is similarly observed in hexagonal MoS$_2$\cite{Mattheiss1973,Mattheiss1973Energy,Kasowski1973,Gan2013,Jin2013,Ganatra2014}. Then, we assume the magnetization axis along the in-plane and run self-consistent calculations till the energy difference converges within 1 $\mu$eV per V atom. Comparing the total energies of the different magnetizations (out-of-plane and in-plane), we find the VSi$_2$N$_4$ has the easy in-plane MA, and the out-of-plane has a higher energy by only 2 $\mu$eV per V atom, which is insufficient to stabilize a 2D Ising magnetization effectively. 

To realize a large orbital moment and giant MA, the single electron in the V$^{4+}$ 3$d^1$ configuration should occupy the $x^2-y^2$/$xy$ doublet, which makes the SOC active and may eventually determine the 2D Ising magnetization. To address this, we perform the GGA+SOC+$U$ calculations and initialize the system in two distinct electronic configurations: one with the single electron occupying the $3z^2-r^2$ state and the other in the $L_{-2}$ state. Our results show that the $L_{-2}$ state is unfavorable in the VSi$_2$N$_4$ monolayer and ultimately transitions to the $3z^2-r^2$ and $x^2-y^2$/$xy$ hybridized orbital state, as visualized in Fig. \ref{eq: 2}. Therefore, the VSi$_2$N$_4$ monolayer has the V$^{4+}$ 3$d^1$ $S$ = 1/2 configuration favoring a hybridized orbital state over the energetically unfavorable $L_{-2}$ state, resulting in a small in-plane MA of 2 $\mu$eV per V atom.

As we mentioned above, the potential for a large orbital moment and huge MA arises from its unique triangular prismatic crystal field. However, this potential is significantly limited due to the strong hybridization between the lower-energy $3z^2-r^2$ orbital and the higher $xy$/$x^2-y^2$ orbitals. In the VSi$_2$N$_4$, each vanadium (V) atom possesses five $d$ orbitals that are initially orthogonal within the triangular prismatic crystal field. Yet, the hybridization occurs due to the influence of adjacent V ions, which themselves have a high coordination number of 6 in the triangular lattice. This results in VSi$_2$N$_4$ exhibiting only a minimal MA.

\begin{figure}[t]
	\includegraphics[width=15cm]{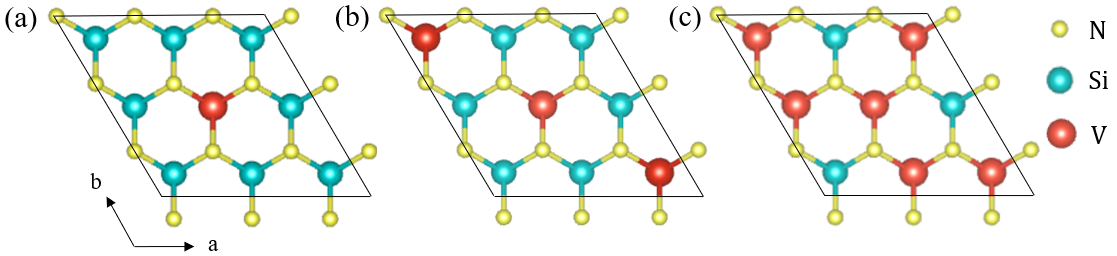}
	\centering
	\caption{Top view into the $ab$ plane of the middle layer VN$_2$ in the $3\times3\times1$ supercell for (a)V$_\frac{1}{9}$Si$_\frac{26}{9}$N$_4$, (b) V$_\frac{1}{3}$Si$_\frac{8}{3}$N$_4$, (c)V$_\frac{2}{3}$Si$_\frac{7}{3}$N$_4$.
	}
	\label{Fig.3}
\end{figure}

\begin{figure}[H]
	\includegraphics[width=8cm]{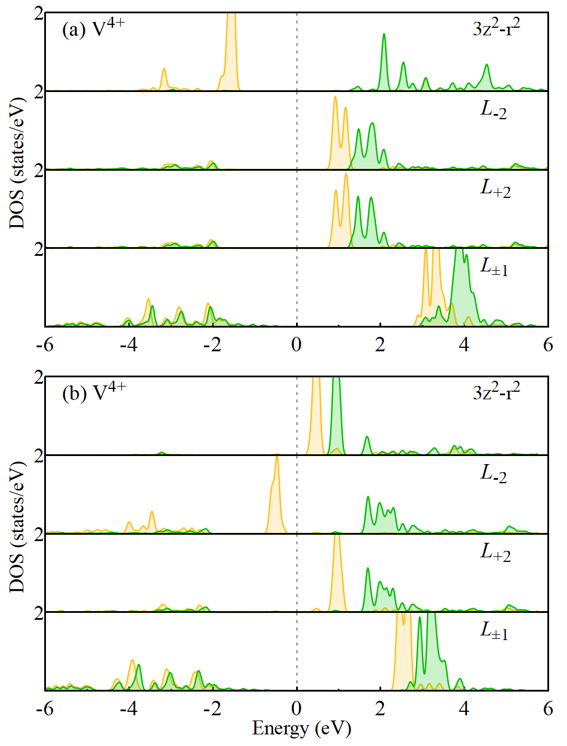}
	\centering
	\caption{(a) $3z^2-r^2$ and (b) $L_{-2}$ states of V$_\frac{1}{9}$Si$_\frac{26}{9}$N$_4$ by GGA+$U$+SOC. The orange (green) curve stands for the up (down) spin. The Fermi level is set at zero energy.
	}
	\label{Fig.4}
\end{figure}

In order to explore the feasibility of achieving the $L_{-2}$ state characterized by a large orbital moment and huge MA, we employ calculations within an isolated triangular prismatic crystal field. For this purpose, a $3\times3\times1$ VSi$_2$N$_4$ supercell comprising 9 V ions is utilized. 
To prevent magnetic interactions, 8 of the V ions are substituted with Si ions named V$_\frac{1}{9}$Si$_\frac{26}{9}$N$_4$, which share the same +4 valence state but are nonmagnetic, as seen in Fig. \ref{Fig.3}(a).
Additionally, Si's conductive band is its 3$p$ orbitals, which cannot be hybridized with the adjacent V 3$d$ orbitals.
Using the corresponding occupation number matrix of the $3z^2-r^2$ and $L_{-2}$ states, our LSDA+SOC+$U$ calculations show that both insulating solutions can be stabilized, as seen in Fig. \ref{Fig.4}.
The $3z^2-r^2$ state is more stable than the $L_{-2}$ state by 756 meV per V atom, agreeing well with expectations based on the isolated triangular prismatic crystal field.
This energy difference is the result of the crystal field energy difference between the $3z^2-r^2$ and the degenerate $xy$ and $x^2-y^2$ orbitals, minus the energy contribution from spin-orbit coupling in the degenerate orbitals. Our calculations demonstrate that the $L_{-2}$ state represents a local minimum in the energy landscape, and we have successfully stabilized this state.
In contrast to the prominent hybridization between the $3z^2-r^2$ and $x^2-y^2$/$xy$ orbitals observed in VSi$_2$N$_4$ (refer to Fig. 2), this interaction is significantly reduced in the isolated V$_\frac{1}{9}$Si$_\frac{26}{9}$N$_4$ composition.

In Fig. \ref{Fig.4}(a), we observe that the single electron of V$^{4+}$ in the 3$d^1$ configuration mainly fills the lowest $3z^2-r^2$ orbital, leaving the higher $x^2-y^2$/$xy$ ($xz$/$yz$) orbitals unoccupied. The calculated magnetic anisotropy energy (MAE), at 59 $\mu$eV per V atom with easy in-plane magnetization, indicates that the $3z^2-r^2$ state is not ideal for sustaining stable 2D magnetism. Conversely, as shown in Fig. \ref{Fig.4}(b), the electron in the V$^{4+}$ ion occupies the higher-energy $x^2-y^2$/$xy$ orbitals when in the $L_{-2}$ state, resulting in a large orbital moment of --1.32 $\mu_{\rm B}$. The calculated MAE value of 23 meV per V atom suggests that the $L_{-2}$ state could support strong Ising-type magnetism.

To enhance magnetic coupling, the coordination number of the V atom is increased in our calculations. We examine two configurations: V$_\frac{1}{3}$Si$_\frac{8}{3}$N$_4$ and V$_\frac{2}{3}$Si$_\frac{7}{3}$N$_4$, as displayed in Figs. \ref{Fig.3}(b) and \ref{Fig.3}(c). Our results show that the V$_\frac{1}{3}$Si$_\frac{8}{3}$N$_4$ structure stabilizes both $3z^2-r^2$ and $L_{-2}$ electronic states. Conversely, in the V$_\frac{2}{3}$Si$_\frac{7}{3}$N$_4$ structure, the $L_{-2}$ state is unstable and converges to the $3z^2-r^2$ state.
Additionally, when examining the V$_\frac{1}{3}$Si$_\frac{8}{3}$N$_4$ structure, the MAE drops to 1 $\mu$eV per V atom for the $3z^2-r^2$ state and to 12 meV per V atom for the $L_{-2}$ state. This reduction is attributed to the interaction between neighboring V ions and is lower than the MAE values found in the V$_\frac{1}{9}$Si$_\frac{26}{9}$N$_4$ configuration.

After careful analysis of the spin-orbital states and Si-substituted levels in the triangular prismatic crystal field of VSi$_2$N$_4$, we find that maximum MAE, large orbital moment, and optimal exchange coupling are attainable in the V$_\frac{1}{3}$Si$_\frac{8}{3}$N$_4$ configuration. However, this configuration exists in a $3z^2-r^2$ ground state, which is contrary to our expectations. To address this, introducing Cr ions with an extra electron could stabilize the desired $L_{-2}$ ground state, which is critical for Ising magnetism.

\begin{figure}[H]
	\includegraphics[width=8cm]{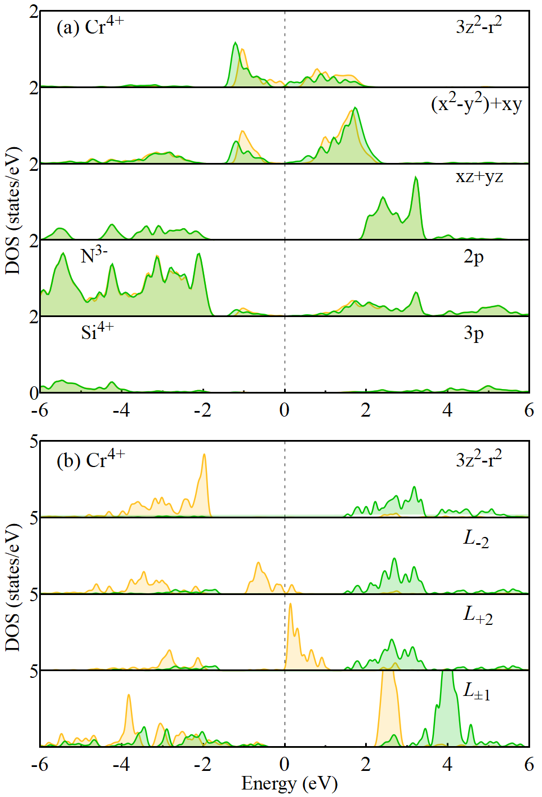}
	\centering
	\caption{(a) The DOS results of the Cr$^{4+}$ 3$d$, N$^{3-}$ 2$p$ and Si$^{4+}$ 3$p$ states for CrSi$_2$N$_4$ by GGA+\textit{U}+SOC. (b) The DOS results of the Cr$^{4+}$ 3$d$ states for Cr$_\frac{1}{3}$Si$_\frac{8}{3}$N$_4$ by GGA+$U$+SOC.
	The orange (green) curve stands for the up (down) spin. The Fermi level is set at zero energy.
	}
	\label{Fig.5}
\end{figure}

We first calculated the pristine CrSi$_2$N$_4$ monolayer. Our results show that it is a non-magnetic material, which has been studied in valley physics. Comparing with the V$^{4+}$ 3$d$$^1$ $S$ = 1/2 state, the extra electron of the Cr$^{4+}$ occupies the hybridization orbitals of the lowest $3z^2-r^2$ singlet and middle $x^2-y^2$/$xy$ doublet, resulting in a nonmagnetic Cr$^{4+}$ with $S$=0 state, as seen in Fig. \ref{Fig.5}(a).

\begin{figure}[t]
	\includegraphics[width=10cm]{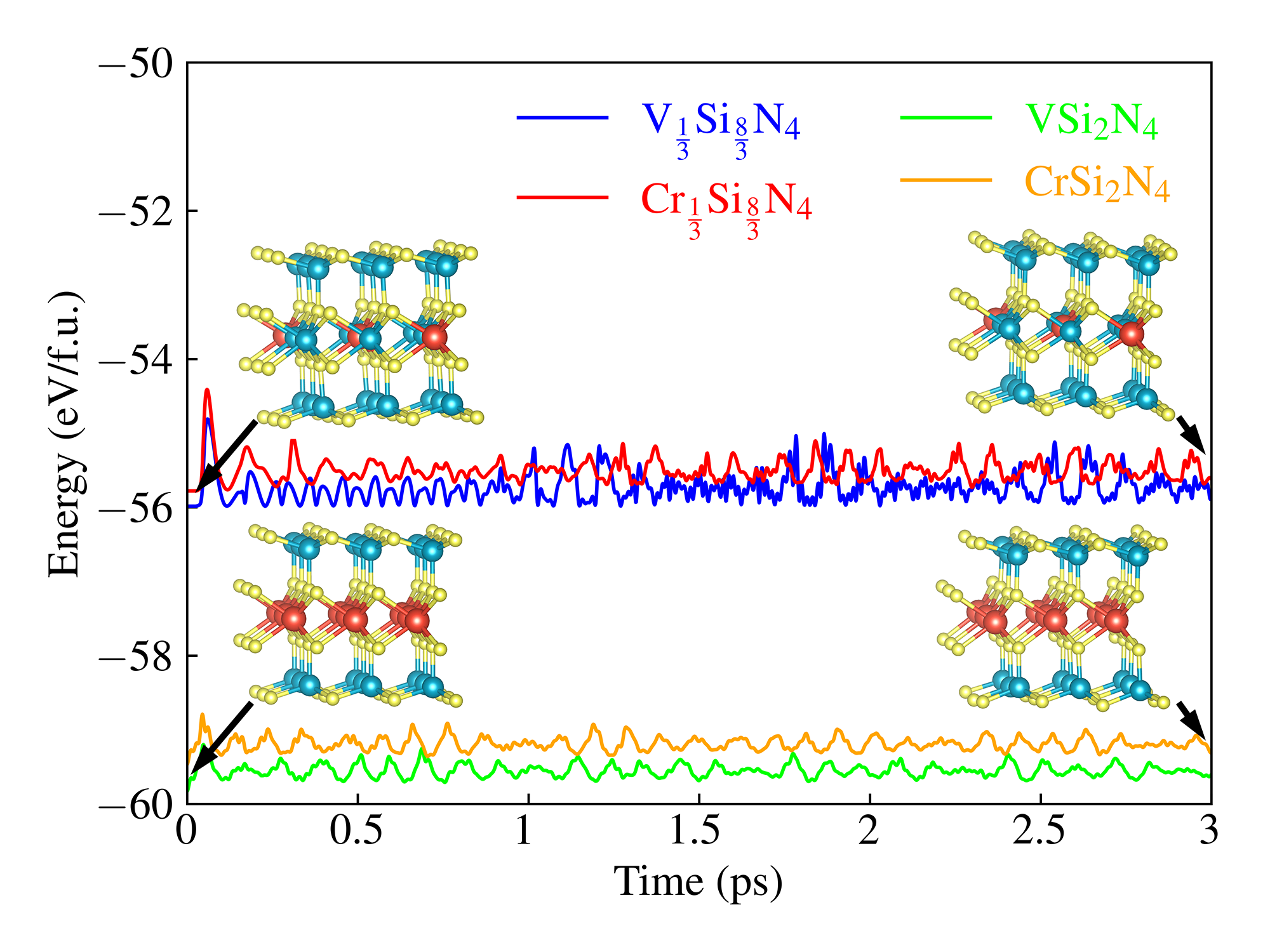}
	\centering
	\caption{Snapshot of the equilibrium structures at 300 K for Cr$_\frac{1}{3}$Si$_\frac{8}{3}$N$_4$, V$_\frac{1}{3}$Si$_\frac{8}{3}$N$_4$,  CrSi$_2$N$_4$ and VSi$_2$N$_4$ at the end of 3 ps of AIMD simulation with a time step of 1.5 fs and the evolution of total energy. 
	}
	\label{Fig.6}
\end{figure}

Then, we calculated the Cr$_\frac{1}{3}$Si$_\frac{8}{3}$N$_4$, a similar structure of Fig. \ref{Fig.3}(b).  The Cr$^{4+}$ exhibits a local spin moment of 2.25 $\mu_{\rm B}$ and large orbital moment of --1.06 $\mu_{\rm B}$. The total spin moment of 6.00 $\mu_{\rm B}$/supercell agrees with the expected Cr$^{4+}$ 3$d^2$ $S$ = 1 spin configuration. 
As seen in Fig. \ref{Fig.5}(b), the lowest $3z^2-r^2$ is fully occupied, while the higher $x^2-y^2$/$xy$ orbitals are half-filled. This results in an active SOC and leads to a nominal Cr$^{4+}$ $(3z^2-r^2)^1(L_{-2})^1$ spin configuration. The presence of sizable orbital moment and the DOS together suggest that  Cr$_\frac{1}{3}$Si$_\frac{8}{3}$N$_4$ has the potential for exhibiting 2D Ising magnetism, making it an attractive candidate for applications in spintronics. To explore the stability of Cr$_\frac{1}{3}$Si$_\frac{8}{3}$N$_4$ under finite temperatures, ab initio molecular dynamics (AIMD) simulations are conducted. Comparisons are also made with CrSi$_2$N$_4$ and VSi$_2$N$_4$. The results indicate that Cr$_\frac{1}{3}$Si$_\frac{8}{3}$N$_4$ remains thermodynamically and dynamically stable at 300 K, as seen in Fig. \ref{Fig.6}.

\renewcommand\arraystretch{1.3}
\begin{table}[t]
	\caption{Relative total energies $\Delta$\textit{E} (meV/supercell), local spin moment ($\mu_{\rm B}$), total spin moment ($\mu_{\rm B}$/supercell) and orbital moment ($\mu_{\rm B}$) of AF and FM states for Cr$_\frac{1}{3}$Si$_\frac{8}{3}$N$_4$ under different strains calculated by GGA+\textit{U}+SOC. ${\parallel}$ (${\perp}$) stands for the in-plane (out-of-plane) magnetization. We use a 3$\times$3$\times$1 supercell of MSi$_2$N$_4$ containing three Cr ions in the Cr$_\frac{1}{3}$Si$_\frac{8}{3}$N$_4$ calculation.
	}
	\begin{tabular}{c@{\hskip2mm}c@{\hskip2mm}c@{\hskip2mm}c@{\hskip2mm}c@{\hskip2mm}c@{\hskip2mm}}
		\hline\hline
		Strain & States & $\Delta$\textit{E} & Cr$_{\rm spin}$ & \textit{M}$_{\rm tot}$  & Cr$_{\rm orb}$   \\ \hline
		{--2.5$\%$}&FM$_{\perp}$     & 17.297 & 2.15       & 6.00  & --0.97       \\
		&AF$_{\perp}$     &     0     & $\pm$2.10  & --2.00 & $\pm$1.10           \\ 
		{   }&FM$_{\parallel}$     & 33.894 (12.194) & 2.16    & 5.99  & --0.93     \\
		&AF$_{\parallel}$      &  21.700 (0) & $\pm$2.10   & --2.00 & $\pm$1.10    \\ \hline
		{0$\%$}&FM$_{\perp}$     & 14.280 & 2.25       & 6.00  & --1.06           \\
		&AF$_{\perp}$     &     0     & $\pm$2.22  & --2.00 & $\pm$1.13            \\ 
		{   }&FM$_{\parallel}$     & 30.800 (10.053) & 2.25    & 5.94  & --0.98      \\
		&AF$_{\parallel}$      &  20.747 (0) & $\pm$2.24   & --2.01 & $\pm$1.11      \\ \hline
		{2.5$\%$}&FM$_{\perp}$     & 9.618 & 2.38       & 6.00  & --1.07            \\
		&AF$_{\perp}$     &     0     & $\pm$2.36  & --2.00 & $\pm$1.10             \\ 
		{   }&FM$_{\parallel}$     & 29.497 (10.149) & 2.39    & 5.99  & --1.02      \\
		&AF$_{\parallel}$      &  19.348 (0) & $\pm$2.36   & --1.99 & $\pm$1.08    \\ \hline\hline
	\end{tabular}
	\label{tb2}
\end{table}

The ground state of Cr$_\frac{1}{3}$Si$_\frac{8}{3}$N$_4$ features a local spin moment of 2.25 $\mu_{\rm B}$ for Cr$^{4+}$, accompanied by an antiparallel orbital moment of --1.06 $\mu_{\rm B}$ along the $z$-axis. This indicates that the SOC, facilitated by the robust SIA, orients the magnetic moment along the $z$-axis, leading to perpendicular MA and giving rise to Ising magnetism. Here we assume the spin Hamiltonian 

\begin{equation} \label{eq: 2}
	\begin{split}
		H=\frac{J}{2}\sum_{i,j}\vec{S_{i}}\cdot \vec{S_{j}}-D\sum_{i}(\vec{S_{i}^{z}})^{2}+{J}'\sum_{i,j}\vec{S_{i}^{z}}\cdot \vec{S_{j}^{z}}
	\end{split}
\end{equation}
where the first term describes the Heisenberg isotropic exchange (AF when $J$\textgreater 0), the second term is the SIA with the easy magnetization $z$-axis (when $D$\textgreater 0), and the last term $J'$ refers to the anisotropic exchange. To determine the magnetic parameters $J$, $D$, and $J'$ in Cr$_\frac{1}{3}$Si$_\frac{8}{3}$N$_4$, we use a 3$\times$3$\times$1 supercell of MSi$_2$N$_4$ containing three Cr ions, as seen in Fig. 3(b). This allows us to flip the spin of the central Cr ion, creating an AF state. However, it is important to note that this AF state can be described as a ferrimagnetic state due to the non-zero net magnetic moment in which two Cr ions have spins aligned in one direction and one in the opposite, leading to a net magnetic moment of --2 $\mu_{\rm B}$ per supercell. We then compute the magnetic properties for four distinct states: FM and AF, each with perpendicular and in-plane magnetization (as detailed in Table \ref{tb2}).
Counting $JS^2$ for each pair of Cr$^{4+}$ $S$ = 1 ion (positive $J$ refers to AF exchange), the magnetic exchange energies of the four states per supercell are written as follows:
\begin{equation} \label{eq: 3}
	\begin{aligned}
		&\ E_{\rm {FM}}^{\perp}=(3J-D+3J')S^{2}&\\
		&\ E_{\rm {AF}}^{\perp}=(-3J-D-3J')S^{2}&\\
		&\ E_{\rm {FM}}^{\parallel}=3JS^{2}&\\
		&\ E_{\rm {AF}}^{\parallel}=-3JS^{2}&\\
	\end{aligned}
\end{equation}

\begin{figure}[t]
	\includegraphics[width=8cm]{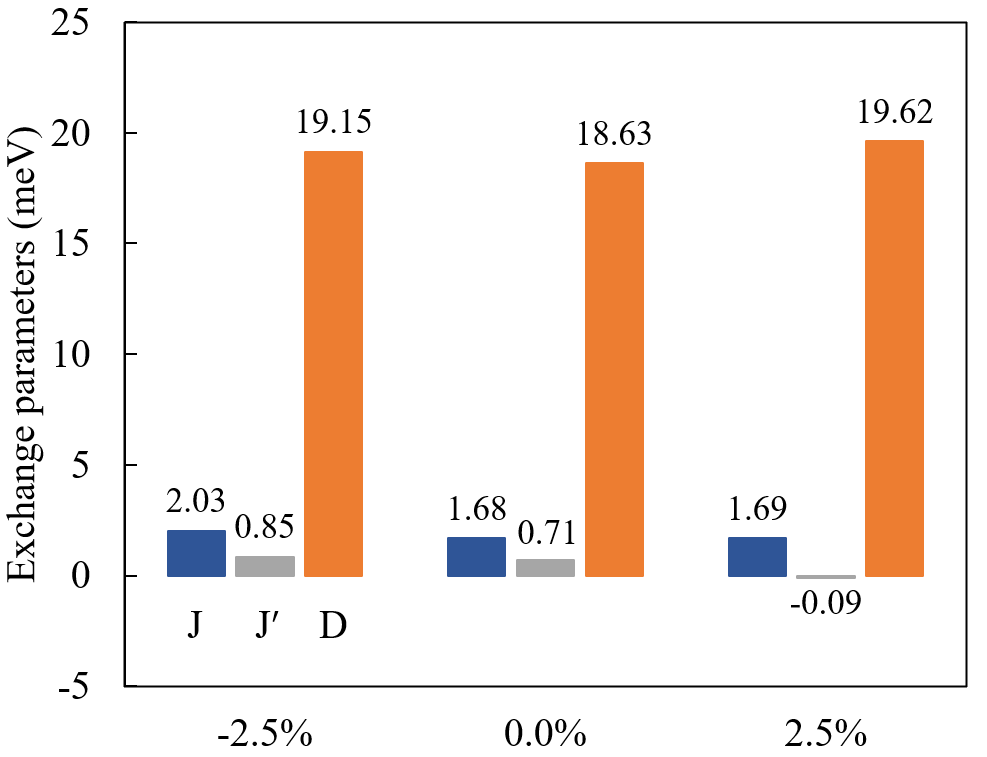}
	\centering
	\caption{The exchange parameters $J$, $J'$ and $D$ (meV) of Cr$_\frac{1}{3}$Si$_\frac{8}{3}$N$_4$ monolayer under different strains.
	}
	\label{Fig.7}
\end{figure}

Based on our total energy calculations, we estimate the magnetic parameters as follows: $J$ = 1.676 meV, $D$ = 18.634 meV, $J'$ = 0.705 meV. One of the key findings is that the Ising-type, represented by the $D$ term, plays a dominant role in establishing perpendicular MA in the Cr$_\frac{1}{3}$Si$_\frac{8}{3}$N$_4$ monolayer.
Specifically, the $D$ is about twenty times stronger than $J'$ in stabilizing the 2D Ising magnetism. The positive $J$ refers to the AF coupling of the adjacent Cr$^{4+}$-Cr$^{4+}$ ions. Therefore, the Cr$_\frac{1}{3}$Si$_\frac{8}{3}$N$_4$ monolayer can be a 2D AF Ising magnetic material. Furthermore, we study a biaxial strain effect on Cr$_\frac{1}{3}$Si$_\frac{8}{3}$N$_4$ monolayer. Our results show that the $(3z^2-r^2)^1(L_{-2})^1$ ground state remains robust against the strains on the optimized lattice. The SIA strength rises in the feasible strain, as seen in Fig. \ref{Fig.7}.

\section{Conclusions}
In summary, we propose Cr$_\frac{1}{3}$Si$_\frac{8}{3}$N$_4$ monolayer can be a 2D AF Ising magnetic material, using crystal field theory, spin-orbital state analyses,  and density functional calculations. Our results indicate that the strong $d$ orbital hybridization between adjacent M$^{4+}$ ions in the MSi$_2$N$_4$ (M = V, Cr) monolayers
disrupts the $d$ orbital splitting in this triangular prismatic crystal field. Through the Si$^{4+}$-substituted, the Cr$_\frac{1}{3}$Si$_\frac{8}{3}$N$_4$ monolayer can achieve the huge perpendicular MA of 18.63 meV per Cr atom with a large orbital moment of --1.06 $\mu_{\rm B}$ along the $z$-axis. 
Our research emphasizes the importance of investigating the degrees of freedom in spin-orbital states as a fruitful avenue for discovering new 2D Ising materials.

\section*{Acknowledgements}
This work was supported by National Natural Science Foundation of China (Grants No. 12104307). S. Chen and W. Xu contributed equally to this work.

\bibliography{VSiN}

\end{document}